# Quenched divergences in the deconfined phase of SU(2) gauge theory

Joe Kiskis

*Lattice Hadron Physics Collaboration*

*Dept. of Physics, University of California, Davis, CA 95616*

Rajamani Narayanan

*American Physical Society, One Research Road, Ridge, NY 11961*

## Abstract

The spectrum of the overlap Dirac operator in the deconfined phase of quenched gauge theory is known to have three parts: exact zeros arising from topology, small nonzero eigenvalues that result in a non-zero chiral condensate, and the dense bulk of the spectrum, which is separated from the small eigenvalues by a gap. In this paper, we focus on the small nonzero eigenvalues in an SU(2) gauge field background at $\beta = 2.4$ and $N_T = 4$. This low-lying spectrum is computed on four different spatial lattices ($12^3$, $14^3$, $16^3$, and $18^3$). As the volume increases, the small eigenvalues become increasingly concentrated near zero in such a way as to strongly suggest that the infinite volume condensate diverges.

## I. INTRODUCTION

In this paper, we will describe a numerical analysis of the small, nonzero eigenvalues of the overlap Dirac operator in the deconfined phase of quenched SU(2) lattice gauge theory.

This is a particularly interesting place to look for singularities associated with the strong infrared divergences that appear in quenched continuum theory in the limit of zero quark mass and infinite volume. As indicated by the Wilson line, this is the deconfined phase, which carries with it a naive expectation of unbroken chiral symmetry. Thus when chiral symmetry breaking effects appear, they are less expected and more dramatic. On the other hand, and somewhat ironically, it may be easier to make a study in this region where the small and important eigenvalues have a relatively low spectral density and are well separated from the bulk of the spectrum. Also it may be easier to study the possible association with instantons since the finite value of $N_T$ leads to relatively sparse and well-separated instantons. We find strong evidence for an infinite volume divergence in the spectral density at zero eigenvalue which implies a divergence in the chiral concentrate. This arises from small *nonzero* eigenvalues that become denser near zero as the volume increases.

Lattice simulations of strong interactions are inevitably first performed in the quenched approximation where all effects from quark loops are ignored. However, there are strong theoretical arguments that quenching the continuum version of the theory produces pathological infrared divergences for massless quarks in infinite volume [1,2]. Thus a correct lattice implementation of massless fermions and chiral symmetry should also have these infrared divergences. On one hand, their appearance is a test for a correct lattice fermion formulation. On the other hand, they must be understood and accounted for in drawing physical conclusions from quenched lattice calculations. Thus without the fermionic action in the path integral, fermionic observables must be interpreted with care. The quark propagator can have anomalous large distance behavior that is associated with singular behavior of the spectral density near zero. This leads to new issues in the discussion of spontaneous chiral symmetry breaking. There are divergent corrections to the usual relation between the quark and pion masses in the broken symmetry phase. This in turn implies that the chiral condensate diverges.

Earlier quenched lattice simulations did not see the effect of these quenched pathologies for a variety of reasons. Corrections due to finite lattice spacing effects turn out to be large for staggered fermions making it difficult to disentangle the quenched divergences. Wilson fermions are plagued by exceptional configurations at small masses in the quenched approximation [3]. The quark propagator diverges due to the zero modes that appear for positive quark masses when chiral symmetry is explicitly broken. In spite of the problems associated with Wilson fermions, some recent simulations of quenched Wilson fermions on larger lattices have shown some evidence for the continuum quenched divergences [4]. Also the use of a modified quenched approximation has given evidence for quenched pathologies [5] predicted by continuum theory. Since Wilson fermions do not allow the exact zero modes due to global topology to be separately identified, the observed signal includes their contribution as a finite volume effect. Studies with several lattice volumes and at fixed lattice coupling could separate the contributions and identify the true quenched divergence present in the thermodynamic limit. Attempts to look for quenched divergences using domain wall fermions [6] have been hampered by larger than expected chiral symmetry breaking from the finite size of the fifth dimension. Overlap Dirac fermions do not have the problems at finite mass and finite volume that come from exceptional configurations and explicitly broken chiral symmetry. In an arbitrary gauge field background, the overlap Dirac propagator can have poles only at zero bare quark mass [7,8]. Therefore there are no singularties in the overlap Dirac propagator at finite volume and non-zero quark mass even in the quenched approximation. First attempts to look for continuum quenched chiral logarithms using the overlap Dirac operator can be found in Ref. [9].

The quenched approximation is also used in finite temperature simulations [10]. Earlier results using staggered fermions seemed to produce a spectrum consistent with an unbroken chiral symmetry in the deconfined phase. Unfortunately, this result is an artifact of the lack of chiral and flavor symmetry for the staggered fermion action on the lattice [11]. The spectrum of the overlap Dirac operator [7] in the deconfined phase does not have a gap [12]. Instead it shows some interesting features. There are three pieces: exact zeros, a low density of very small eigenvalues, and the dense part of the spectrum, which is separated from zero by a gap. Furthermore the very small eigenvalues appear to have a finite spectral density at zero and thus produce a non-zero chiral condensate. We should emphasize here that this is the spectrum of non-zero eigenvalues obtained after one has removed all the exact zero eigenvalues of the overlap Dirac operator due to global topology of the gauge fields.

## II. CALCULATIONS AND RESULTS

Our background fields were SU(2) gauge fields generated on an $L^3 \times 4$ lattice at $\beta = 2.4$ using the standard Wilson action. All configurations were forced to have a positive expectation value for the Wilson line and anti-periodic boundary conditions were imposed on the fermions in the time direction. This eliminates an "unphysical" signal for a chiral condensate [13]. To study the thermodynamic limit, we generated gauge field configurations on four different lattices listed in Table I. We computed the spectrum of the square of the massless hermitian overlap Dirac operator,

$$H_o = \frac{1}{2}(\gamma_5 + \epsilon(H_w(m_w))); \quad [\gamma_5, H_o^2] = 0 \qquad (1)$$

where the argument of $\epsilon$ is the standard hermitian Wilson Dirac operator with the mass $m_w$ set to 1.5[1]. The calculations were embarrassingly parallel and were carried out on PC's with Pentium II, Pentium III, and Athlon processors. The computation was performed in double precision for a reasonably accurate determination of the small eigenvalues. The polar approximation [14] for the $\epsilon$ function was used. Small eigenvalues of $H_w$ were projected out before the action of $\epsilon$ and were treated exactly. In practice the action of $\epsilon$ on a vector was obtained to a very high precision and the "two–pass" algorithm [15] was used to minimize memory requirements. Based on previous work in the deconfined phase [12], we used 0.05 as the cutoff for the small eigenvalues, and all eigenvalues below 0.05 were computed using the Ritz algorithm [16]. The bulk of the spectrum becomes dense above roughly 0.2. There are some loners in the desert that lies between 0.05 and 0.2. We computed the eigenvalues in both chiral sectors to unambiguously separate the exact zero eigenvalues from the non-zero ones. The combined average number of exact zeros and small, nonzero eigenvalues per unit volume is $\langle n \rangle$. The total number of pairs of small, nonzero eigenvalues in each ensemble is $n_s$. These data are listed in Table I.

---

[1] Our convention is such that this mass corresponds to negative quark mass for Wilson fermions at this lattice coupling.

| $L$ | $N$ | $\langle n \rangle$ | $n_s$ |
|---|---|---|---|
| 12 | 1014 | $7.7(3) \times 10^{-5}$ | 58 |
| 14 | 538 | $8.0(4) \times 10^{-5}$ | 70 |
| 16 | 286 | $7.7(4) \times 10^{-5}$ | 66 |
| 18 | 146 | $8.5(5) \times 10^{-5}$ | 72 |

TABLE I. $L$ is the spatial extent of the lattice, $N$ is the number of configurations generated, $\langle n \rangle$ is the average number of small eigenvalues including exact zero modes per unit volume and $n_s$ is the total number of pairs of small, non-zero eigenvalues in each ensemble

Since one expects gauge fields of all topology even in the deconfined phase, instantons and anti-instantons are present. But the topological susceptibility is small in the deconfined phase, and the gas of instantons and anti-instantons is dilute. The four values for $\langle n \rangle$ in Table I are equal within errors. This is consistent with the hypothesis that the small eigenvalues are associated with a dilute gas of approximately noninteracting instantons and anti-instantons and that this gas has a good thermodynamic limit. This is not a surprise since we do not expect anything pathological about the gauge field configurations themselves. Including the exact zero eigenvalues in each configuration, the spectrum of $H_o^2$ has $n_+$ small eigenvalues of positive chirality and $n_-$ eigenvalues of negative chirality. Assuming $n_+$ is greater than $n_-$, we will have $Q = n_+ - n_-$ exact zeros of positive chirality and $n_-$ non-zero eigenvalues of each chirality, which are paired. As in the previous work [12], we find that the probability distribution, for $n_+$ and $n_-$, is consistent with that of $n = n_+ + n_-$ non-interacting objects. When combined with a numerical study of the fermion spectrum in a background of instantons and anti-instantons [17], this picture also suggests that the chiral condensate diverges.

Let $\rho(E)$ be the density of non-zero eigenvalues per unit energy and per unit volume. The bare chiral condensate in the continuum is given by

$$\bar{\psi}\psi(M) = \int_0^\infty dE \frac{2M\rho(E)}{E^2 + M^2} \tag{2}$$

If $\rho(E)$ goes to zero as $E$ goes to zero, then there is no condensate. If $\rho(E)$ is finite as $E$ goes to zero, then we have a finite condensate. If $\rho(E)$ diverges as $E$ goes to zero, we have an infinite condensate. Since numerical computations are performed on a finite lattice, the spectrum per configuration is discrete, and it is convenient to use the cumulative quantity $N(E, V)$ which is the number of nonzero eigenvalues below $E$ in a lattice of volume $V = L^3 \times N_T$ averaged over all configurations. It is from this that $\rho$ is defined

$$\rho(E) = \frac{d}{dE} \lim_{V \to \infty} \frac{N(E, V)}{V}. \tag{3}$$

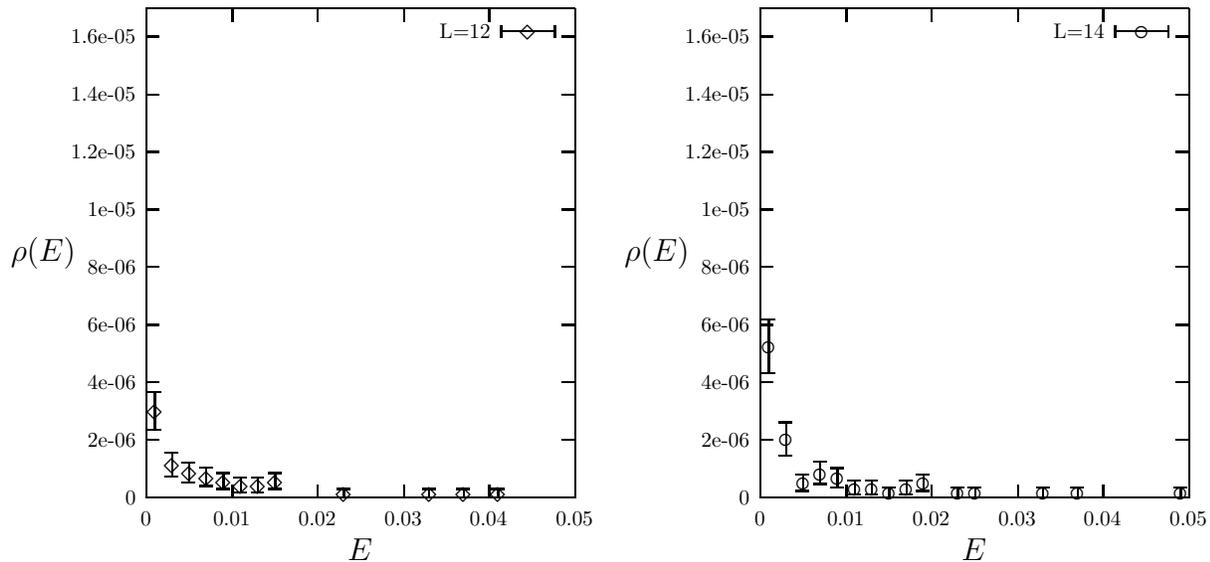

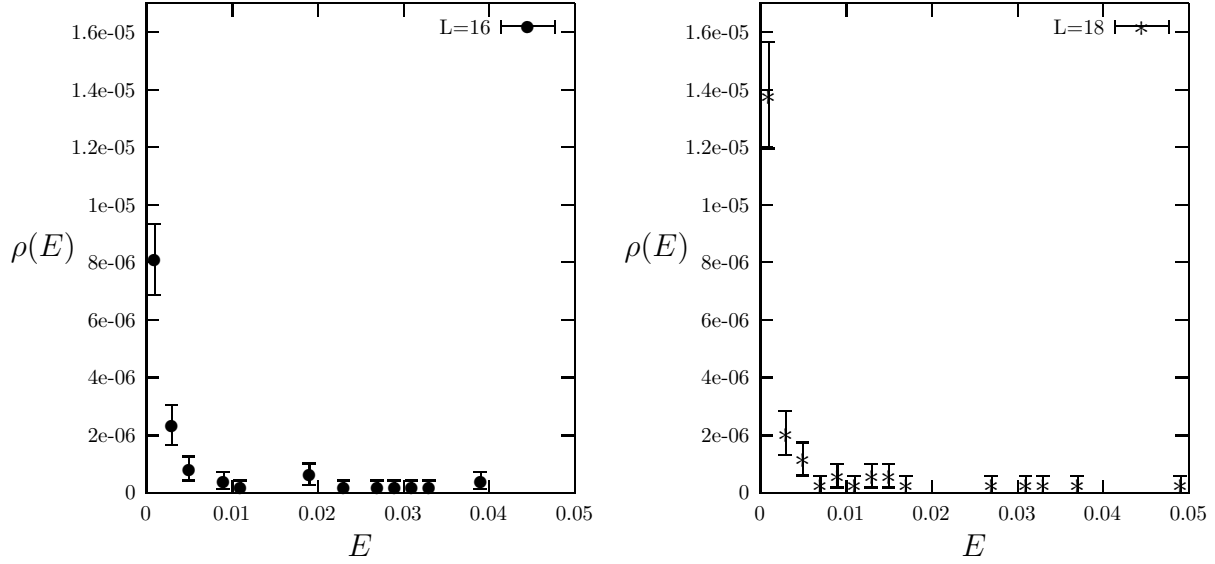

Figure 1: Distribution of the small eigenvalues per unit volume on four different lattices.

A qualitative inspection of the spectral densities of the non-zero eigenvalues in Figure 1 shows that the spectrum becomes more sharply peaked near zero as the volume increases. An alternative presentation of the data is $N(E,V)/(EV)$ vs. $E$ in Figure 2. A divergence in $N(E,V)/(EV)$ as $E$ goes to zero leads to a divergent condensate. Based on this range of volumes, a divergence for $V \to \infty$ followed by $E \to 0$ is indicated. A least square fit of the data on the $18^3 \times 4$ lattice yields

$$\frac{N(E,V)}{EV}\Big|_{L=18} \approx 4.2 \cdot 10^{-5} \times E^{-0.80}. \tag{4}$$

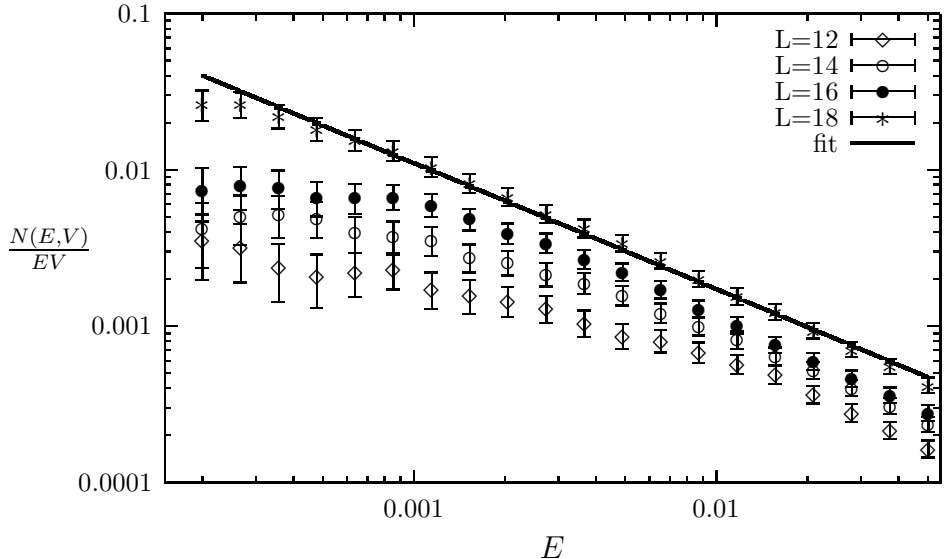

Figure 2: A plot of $N(E,V)/(EV)$ from the small nonzero eigenvalues on four different lattices plotted on a log-log scale. The solid line is a least squares fit to the $L = 18$ data weighted by the statistical errors.

We can also directly compute the contribution of the small, nonzero eigenvalues to the chiral condensate [8]

$$\langle \bar{\psi}\psi(M) \rangle = \langle \frac{1}{V} \sum_{E>0} \frac{2M(1-E^2)}{E^2(1-M^2)+M^2} \rangle. \qquad (5)$$

The contribution from the bulk of the spectrum is not relevant at small masses where it gives a term linear in $M$. The result is plotted in Figure 3 for the four different lattices. Again, the curves do not seem to have an infinite volume limit in the small mass region. This can be used to test the functional form found in [1] where a small negative power of $M$ is predicted. A least square fit of the $L = 18$ curve weighted by the statistical errors gives

$$\langle \bar{\psi}\psi(M) \rangle \Big|_{L=18} \approx 8.1 \cdot 10^{-5} \times M^{-0.81}. \qquad (6)$$

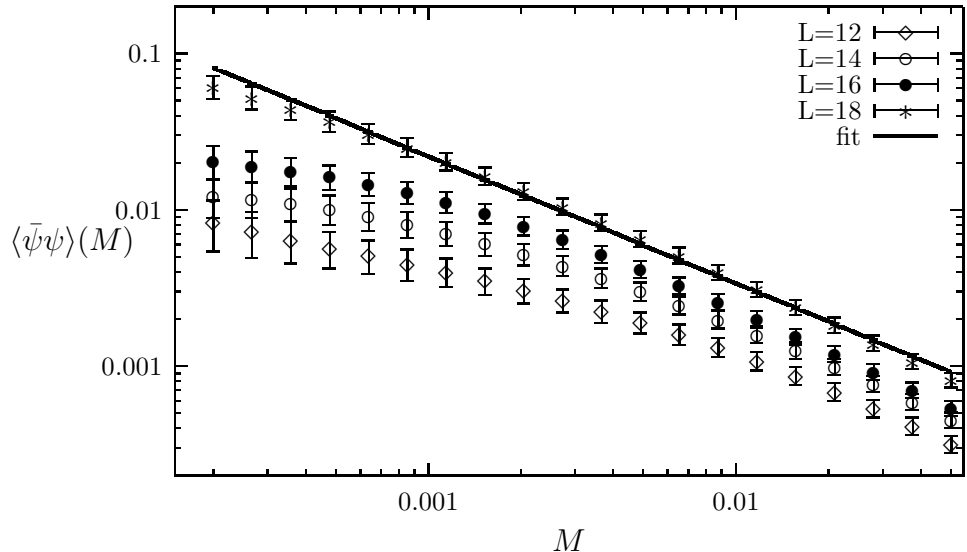

Figure 3: Contribution to the chiral condensate from the small nonzero eigenvalues on four different lattices plotted on a log-log scale. The solid line is a least squares fit to the $L = 18$ data weighted by the statistical errors.

This fit is consistent with the fit of $N(E,V)/(EV)$ in Figure 2. These fits are entirely based on the $18^3 \times 4$ data and do not necessarily imply that the same form or power would work at larger volume. The power is far more negative than the zero-temperature estimates [1] and is simply a reflection of the strong peaking of the spectrum with increasing volume seen in Figure 2. The strong divergence could arise from the dilute gas of instantons. If the gas is dilute, the would-be zero modes will stay very close to zero.

With the assumption that there are Goldstone bosons in this region, we can view this result from the perspective of the associated effective theory based on a chiral lagrangian. As a consequence of the finite extent in imaginary time, the leading infrared structure will be three dimensional, and the infrared divergences of the quenched chiral perturbation theory will be stronger than was the case for zero temperature. Thus power divergences can arise in single loop diagrams rather than as an infinite sum of logs, and could explain the stronger divergence at finite temperature.

## III. CLOSING COMMENTS

Following this investigation of quenched divergences in the quark spectrum, it would also be interesting to look at meson correlators at small masses [18]. A loss of positivity is predicted in the scalar meson propagator. The scalar susceptibility is the derivative of the chiral condensate in the quenched approximation [8]. Since the divergent part of the chiral condensate should dominate in the massless limit, the scalar susceptibility should be negative there. But it is necessary to remove the contribution from exact zero modes since these also cause a loss of positivity [19]. In the near future, we plan to the study the contribution of the small modes to meson correlators in the deconfined phase. For other future work, studies with varying lattice gauge coupling and Wilson mass $m_w$ (Equation 1) should be done.

In this paper, we have shown numerical evidence for a quenched divergence in the chiral condensate for SU(2) gauge theory in the deconfined phase. The divergence is caused by a finite density of small eigenvalues whose distribution becomes increasingly peaked near zero as the thermodynamic limit is approached. We should emphasize again that this divergence is due to the nonzero eigenvalues of the overlap Dirac operator. The divergence from the exact zeros, which we have not included, is a finite volume effect that does not persist in the thermodynamic limit.

## ACKNOWLEDGMENTS

R.N. would like to acknowledge the DOE grant DE-FG02-01ER41165 for computational support. R.N. would like to thank the Physics Department at BNL for the use of the BNL computing facility. J.K. gratefully acknowledges access to the linux PC cluster of W. Pickett's condensed matter theory group at UC Davis where some of the calculations were done.